\documentclass{sigplanconf}

\usepackage{amsmath,listings,url,soul,color}

\begin{document}

\setlength{\pdfpageheight}{\paperheight}
\setlength{\pdfpagewidth}{\paperwidth}

\conferenceinfo{IFL '16}{Aug 31- Sep 2, 2016, Leuven, Belgium} 
\copyrightyear{2016} 
\copyrightdata{978-1-nnnn-nnnn-n/yy/mm} 
\doi{nnnnnnn.nnnnnnn}




\titlebanner{banner above paper title}        
\preprintfooter{short description of paper}   

\title{Embedded SML using the MLton compiler}

\authorinfo{Jeffrey Murphy}
           {University at Buffalo}
           {jcmurphy@buffalo.edu}
\authorinfo{Bhargav Shivkumar}
           {University at Buffalo}
           {bhargavs@buffalo.edu}
\authorinfo{Lukasz Ziarek}
           {University at Buffalo}
           {lziarek@buffalo.edu}

\maketitle

\begin{abstract}
In this extended abstract we present our current work on leveraging
Standard ML for developing embedded and real-time systems. Specifically
we detail our experiences in modifying MLton, a whole program, optimizing
compiler for Standard ML, for use in such contexts.  We focus primarily on
the language runtime, re-working the threading subsystem and garbage collector,
as well as necessary changes for integrating MLton generated programs
into a light weight operating system
kernel.  We compare and contrast
these changes to our previous work on extending MLton for multicore systems, which focused around
acheiving scalability.
\end{abstract}

%
%

\section{Introduction}
\label{sec:intro}
With the resurgence of popularity of functional programming,
practitioners have begun re-examining usage of functional programming languages
for embedded and real-time systems~\cite{Li:2016:ARC:2930957.2930958,Hudak1999,JFP:9060502,Hammond2001,Wan:2001:RF:507635.507654,DBLP:conf/sfp/Hammond03}
Recent advances in program verification~\cite{Audebaud2009568,Almeida:2013:CCC:2541806.2516652,Kumar:2014:CVI:2535838.2535841} and formal methods~\cite{Arts:2004:DVE:1010992.1010995,Lopez:2002:SPA:647983.743555,Sherrell:1994:ETZ:213488.213492}
make functional programming
languages appealing, as embedded and real-time systems have more stringent correctness criteria.
Since many embedded boards are now multi-core, advances in parallel and concurrent programming
models and language implementations for  functional languages are also appealing as lack of mutable
state often results in simpler reasoning about concurrency and parallelism.

Many functional language runtimes, however, do not have support needed to deploy on embedded boards.  Although most
function languages provide efficient implementations, their compiler optimizations, runtime support, and programming
models are geared toward general purpose desktop applications~\cite{acml} or focus on scalability in distributed systems~\cite{Erlang}
or on large parallel machines~\cite{JFP:9547587,spoonhower2009scheduling}.  
We use MLton~\cite{MLton}, a whole program optimizing compiler for SML, as a base to implement the constructs necessary for using Standard ML (SML)~\cite{Milner:1997:DSM:549659} in an embedded context. We introduce a native threading model built on top of POSIX threads and a mechanism to induce priorities among these threads. Our system supports running programs built using this system on RTEMS, a real-time operating system. In this paper we describe our experience from these efforts and
compare and contrast our previous design for MultiMLton~\cite{JFP:9547587}.


\section{Threaded programming in MLton}
\label{sec:background}
MLton is an open-source, whole-program, optimizing SML compiler that generates very efficient executables. MLton has a number of features that are well suited for embedded systems.  MLton uses aggressive whole program optimizations for improving both performance as well as memory utilization. MLton provides the capability of generating c-code, making cross compilation for a different architecture other than the host architecture relatively easy. MLton provides a concurrent, but not parallel, threading model with support for communication between threads either over shared memory of through message passing abstractions.

MLton created threads are green threads that are multiplexed over a single OS level thread. MLton's thread API is well suited for implementing user defined schedulers, including preemptive and cooperative threading models as well as Concurrent ML~\cite{cml} and Asynchronous CML~\cite{acml}. A thread in MLton is a lightweight data structure that represents a paused computation. Threads contain the currently saved execution state of the program, namely the call stack. When a thread is paused, a copy of its current stack is saved and when it is switched to, the stack is restored. MLton also provides a logical ready queue from which the next runnable thread is accessed by the scheduler. This is a regular FIFO queue with no notion of priority, however the structure is implicit, relying on continuation chaining and is embedded in the thread switching code itself. This means that there is no single data  structure that governs threads nor is there an explicit scheduler.
Threading and concurrency libraries (e.g. CML and ACML) built on top of the MLton threading primitives, therefore,
typically introduce their own threading primitives, scheduler, policy, as well as structures for managing ready, suspended,
and blocked threads.


\section{Adapting for OS Threads}
\label{sec:threadingchanges}
Most embedded boards today have more than one core and this might make highly scalable implementations like MultiMLton a preferred choice for embedded use. Such systems, more often than not, go with thread local heaps which have additional read/write barriers to deal with and  overheads associated with global synchronization for a global GC. Although MultiMLton uses techniques such as procrastination (delaying writes that would cause eviction to the global heap) and cleanliness (copying of mutable state to avoid sharing of state)~\cite{Sivaramakrishnan:2012:ERB} to reduce some of these overheads, the true potential of these multicore optimizations are evident when the underlying architecture has 16 or more cores. On the contrary systems like MLton are highly optimized to run on single core machines and we would like to utilize this know how to build a shared heap implementation and give it the ability to spread out across more than one core if needed. This model, we feel, is closer to the current state of embedded boards. 

The first step to having a threading model that supports OS-level concurrency is to split the single thread of execution over multiple OS-level threads. We need to identify and separate the execution states of each of the OS-level threads. MLton keeps track of the state of the system using the \texttt{GC\_STATE} structure. This structure has numerous fields that store the current position of frontier, current executing green thread, current StackTop/StackBottom among others and all these values are accessed at any time by offsetting a pointer to this structure. The decision to use one single structure for storing all the global state was to make the access really fast by caching the entire structure on a register. When there is a single thread of execution, there is no need to worry about concurrent access to the \texttt{GC\_STATE} and thus the integrity of the state is maintained. Introducing multiple threads of execution brings in a plethora of changes including the necessity to differentiate between the thread of execution to which the value being stored belongs. For example, the \texttt{GC\_STATE} must now store the StackTop of all the executing OS threads. Needless to say, threads must also have controlled access to the shared fields in this structure.


As we alluded above, MLton has a concept of green threads, and we are adding OS threads. The \texttt{GC\_thread} structure includes a pointer to the thread's stack. When MLton's thread library wants to switch to a new thread, it calls into the runtime and the pointer to the thread structure is copied into \texttt{GC\_STATE->currentThread} and the \texttt{GC\_STATE} stack related fields are updated with values derived from \texttt{GC\_thread->stack}. Execution then can resume in the newly restored green thread. 

In MultiMLton, our previous design centered around making \texttt{GC\_STATE} an array, with each array element corresponding to a POSIX thread pinned to a processor core. Multiple \texttt{GC\_STATE}s are imperative to MultiMLton as each processor core has its own local heap, and it would be faster to access the execution state of the thread pinned to that core if it were cached in a register local to that core. Any shared state could be relayed using messages between the threads. In RTMLton, we've decided to keep \texttt{GC\_STATE} as a single structure, but implement arrays within it where appropriate. This allows us to be somewhat more optimal when it comes to memory utilization -- an important consideration when targeting embedded systems. For example, in MultiMLton, finding the current green thread running within the OS thread, we would refer to the index \texttt{GC\_STATE[corenumber]->currentThread} but in RTMLton we refer to \texttt{GC\_STATE->currentThread[osthreadnumber]}. This may seem like a trivial distinction, but it allows us to avoid duplicating \texttt{GC\_STATE} fields that do not need to be copied across OS threads, for example various global configuration fields as well as GC configuration fields.

\subsection{Concurrent GC}
To implement concurrent GC, it is necessary to have the garbage collector on its own thread so that it can work independently. But this is a difficult task as in MLton, the GC is very tightly coupled with the main thread of execution. Not only does MLton insert pre-calculated GC checkpoints in the compiled code, it also uses the GC to grow the SML call stack.	Multi-core implementations of SML like Multi-MLton take a different route in handling this separation. They use a per thread heap and thus have a per thread GC which stays coupled to the execution thread. Multiple heaps may pose other complexities (like read/write barrier overheads, global synchronization) in a multi-threaded system, which is why we seek to implement this system on a shared heap. A shared heap implementation is easier but brings us back to the difficult task of pulling out the GC onto a separate thread. In doing so, we need to make sure each thread is responsible for growing its own stack and only invokes the GC when it needs more space on the heap.

As a result of having the garbage collector on a separate thread and by virtue of this implementation being on a single shared heap, we need to ensure that the other threads in the system do not interfere with the GC when it is either scanning or collecting the heap. Other threads working on the heap must be paused before allowing the GC to run but we must also ensure that it is safe to pause these threads when we do so. This is necessary because MLton stores temporary variables on the stack and if the GC were to run before the stack frame is fully reified, the results would be unpredictable. MLton also will write into a newly created stack frame before finalizing and recording the size of the frame. Without the identification of safe points to pause the threads, the heap will be in an inconsistent state that is not conducive to a GC.


Fortunately, MLton identifies these safe points for us, namely GC safe points. GC safe points in MLton are points in code where it is safe for the thread running the code to pause allowing the GC to run. We can use these safe points in our multi-threaded system to identify points where it can be safe for us to pause the threads and thereby allow the GC thread to continue its execution.

Although GC safe points are pre-identified for us, the code generated by the compiler assumes a single threaded model and so we found problematic constructs such as global variables and reliance on caching important pointers in registers for performance. In order to move MLton to a multi-threaded architecture, we needed to rework these architectural decisions. As discussed above, MLton tracks a considerable amount of global state using the \texttt{GC\_STATE} structure so we must refactor this structure, in particular, to make it thread-aware. MLton also uses additional global state, outside of \texttt{GC\_STATE} structure, to implement critical functionality.


For example, MLton caches the frontier pointer in a register. This naturally leads to unpredictable behavior on multi-core as multiple threads attempt to move the frontier during allocations. In terms of global variables, MLton emits C code that mimics machine code and registers. For example, the C code in Listing~\ref{lst:code1} shows MLton comparing a value in the global GC state to the top of the current stack. The result of that comparison is temporarily saved to a global C variable (via the G macro) until it can be operated on by the BNZ macro. Listing~\ref{lst:code2} shows how the G macro expands to reference a globally declared array. In a multi-threaded environment, we must address the use of global variables in these instances, or the compiler must identify these instructions as a critical section and place a barrier around them. We elected to accept the higher memory cost of per-thread global variables versus the higher performance cost of barriers. Listing~\ref{lst:code3} shows how we have the compiler add another dimension to the array, allowing us to isolate the global to a specific thread (where TID represents an zero-relative integer ID unique to each thread).

\lstset{language=C}\begin{lstlisting}[caption={Use of global variables},label={lst:code1}]
    G(Word32, 0) = CPointer_lt 
       (O(CPointer, GCState, 40), StackTop);
    BNZ (G(Word32, 0), L_8);
\end{lstlisting}

\lstset{language=C}\begin{lstlisting}[caption={Expansion of global variables},label={lst:code2}]
    GlobalWord32[0] = CPointer_lt 
       (O(CPointer, GCState, 40), StackTop);
    BNZ (GlobalWord32[0], L_8);
\end{lstlisting}
 
\lstset{language=C}\begin{lstlisting}[caption={MT tolerant global variables},label={lst:code3}]
    GlobalWord32[TID][0] = CPointer_lt 
       (O(CPointer, GCState, 40), StackTop);
    BNZ (GlobalWord32[TID][0], L_8);
\end{lstlisting}


\section{RTEMS support}
\label{sec:rtems}
RTEMS is a real-time operating system that can be run in an emulator, like QEMU, on commodity hardware such as a laptop, or on purpose-built hardware such as a LEON3 board. Since MLton is able to emit C code, it was natural to attempt to cross-compile that code using RTEMS.  

An interesting feature of RTEMS is that resources such as memory size, maximum stack size, number of threads, device driver support, and so on, are specified at compile time. This is interesting because it directly supports the construction of predictable software by specifying the precise characteristics of the underlying OS. The compiler will then produce a bootable binary file that contains the embedded application as well as the OS. An interesting area of research that we are pursuing is tighter integration of the compiler optimization passes with RTEMS. For example, by enhancing the Static Single Assignment (SSA) optimization pass to recognize things such as maximum thread utilization or worst case memory utilization, we can drive better automation of the RTEMS build process.

Given MLton's flexible integration with an installed C-compiler, we were able to modify MLton's build environment to be aware of the availability of the RTEMS cross compiler. We then needed to identify system calls that were not available in RTEMS so that we could eliminate them from the emitted C code or substitute alternatives. This included some memory management calls such as \texttt{memunmap} as well as all networking specific functions. Next we needed to deal with MLton's auto-identification of system constants. Similar to \texttt{autoconf}, MLton will probe the system for various system-specific constants (such as IO constants, POSIX return codes, etc) that can then be referenced from SML code. However, two challenges occurred. First, RTEMS did not provide any networking modules, so the constants related to socket operations need to be suppressed. The second challenge is that the constants identification program needs to be compiled for RTEMS and run within a emulator in order to capture all of the system specific constants which are then used in later stages of the MLton build process. We ran the constants gathering stage of the compilation in QEMU and then stored the output to a file. During the build process, we skip the constants stage and instead copy our RTEMS specific file into the build tree.



\acks

This research is primarily funded by the National Science Foundation, under grant number
CNS-1405614.

\bibliographystyle{plain}

\bibliography{all}

\begin{thebibliography}{10}

\bibitem{Almeida:2013:CCC:2541806.2516652}
Jos{\'e}~Bacelar Almeida, Manuel Barbosa, Gilles Barthe, and Fran\c{c}ois
  Dupressoir.
\newblock Certified computer-aided cryptography: efficient provably secure
  machine code from high-level implementations.
\newblock In {\em Proceedings of the 2013 ACM SIGSAC conference on Computer
  \&\#38; communications security}, CCS '13, pages 1217--1230, New York, NY,
  USA, 2013. ACM.

\bibitem{Arts:2004:DVE:1010992.1010995}
Thomas Arts, Clara Benac~Earle, and John Derrick.
\newblock Development of a verified erlang program for resource locking.
\newblock {\em Int. J. Softw. Tools Technol. Transf.}, 5(2):205--220, March
  2004.

\bibitem{Audebaud2009568}
Philippe Audebaud and Christine Paulin-Mohring.
\newblock Proofs of randomized algorithms in coq.
\newblock {\em Science of Computer Programming}, 74(8):568 -- 589, 2009.
\newblock Special Issue on Mathematics of Program Construction (MPC 2006).

\bibitem{JFP:9060502}
EDWIN BRADY.
\newblock Idris, a general-purpose dependently typed programming language:
  Design and implementation.
\newblock {\em Journal of Functional Programming}, 23:552--593, 9 2013.

\bibitem{Erlang}
Erlang programming language official website.
\newblock \url{http://www.erlang.org/}.

\bibitem{Hammond2001}
Kevin Hammond.
\newblock {\em Implementation of Functional Languages: 12th Int'l Workshop, IFL
  2000 Aachen, Germany, September 4--7, 2000 Selected Papers}, chapter The
  Dynamic Properties of Hume: A Functionally-Based Concurrent Language with
  Bounded Time and Space Behaviour, pages 122--139.
\newblock Springer Berlin Heidelberg, Berlin, Heidelberg, 2001.

\bibitem{DBLP:conf/sfp/Hammond03}
Kevin Hammond.
\newblock Is it time for real-time functional programming?
\newblock In Stephen Gilmore, editor, {\em Revised Selected Papers from the
  Fourth Symposium on Trends in Functional Programming, {TFP} 2003, Edinburgh,
  United Kingdom, 11-12 September 2003.}, volume~4 of {\em Trends in Functional
  Programming}, pages 1--18. Intellect, 2003.

\bibitem{Hudak1999}
Paul Hudak.
\newblock {\em Functional Reactive Programming}, pages 1--1.
\newblock Springer Berlin Heidelberg, Berlin, Heidelberg, 1999.

\bibitem{Kumar:2014:CVI:2535838.2535841}
Ramana Kumar, Magnus~O. Myreen, Michael Norrish, and Scott Owens.
\newblock Cakeml: A verified implementation of ml.
\newblock In {\em Proceedings of the 41st ACM SIGPLAN-SIGACT Symposium on
  Principles of Programming Languages}, POPL '14, pages 179--191, New York, NY,
  USA, 2014. ACM.

\bibitem{Li:2016:ARC:2930957.2930958}
Muyuan Li, Daniel~E. McArdle, Jeffrey~C. Murphy, Bhargav Shivkumar, and Lukasz
  Ziarek.
\newblock Adding real-time capabilities to a sml compiler.
\newblock {\em SIGBED Rev.}, 13(2):8--13, April 2016.

\bibitem{Lopez:2002:SPA:647983.743555}
Natalia L\'{o}pez, Manuel N\'{u}\~{n}ez, and Fernando Rubio.
\newblock Stochastic process algebras meet eden.
\newblock In {\em Proceedings of the Third International Conference on
  Integrated Formal Methods}, IFM '02, pages 29--48, London, UK, UK, 2002.
  Springer-Verlag.

\bibitem{Milner:1997:DSM:549659}
Robin Milner, Mads Tofte, and David Macqueen.
\newblock {\em The Definition of Standard ML}.
\newblock MIT Press, Cambridge, MA, USA, 1997.

\bibitem{MLton}
{ML}ton.
\newblock \url{http://www.mlton.org}.

\bibitem{cml}
John~H. Reppy.
\newblock {\em Concurrent Programming in ML}.
\newblock Cambridge University Press, New York, NY, USA, 1999.

\bibitem{Sherrell:1994:ETZ:213488.213492}
Linda~B. Sherrell and Doris~L. Carver.
\newblock Experiences in translating z designs to haskell implementations.
\newblock {\em Softw. Pract. Exper.}, 24(12):1159--1178, December 1994.

\bibitem{JFP:9547587}
K.~C. Sivaramakrishnan, Lukasz Ziarek, and Suresh Jagannathan.
\newblock {MultiMLton}: {A} multicore-aware runtime for standard {ML}.
\newblock {\em Journal of Functional Programming}, 24:613--674, 2014.

\bibitem{Sivaramakrishnan:2012:ERB}
KC~Sivaramakrishnan, Lukasz Ziarek, and Suresh Jagannathan.
\newblock Eliminating read barriers through procrastination and cleanliness.
\newblock In {\em Proceedings of the 2012 Int'l symposium on Memory
  Management}, ISMM '12, pages 49--60, New York, NY, USA, 2012. ACM.

\bibitem{spoonhower2009scheduling}
Daniel~John Spoonhower.
\newblock {\em Scheduling deterministric parallel programs}.
\newblock ProQuest, 2009.

\bibitem{Wan:2001:RF:507635.507654}
Zhanyong Wan, Walid Taha, and Paul Hudak.
\newblock Real-time frp.
\newblock In {\em Proceedings of the Sixth ACM SIGPLAN Int'l Conf. on
  Functional Programming}, ICFP '01, pages 146--156, New York, NY, USA, 2001.
  ACM.

\bibitem{acml}
Lukasz Ziarek, KC~Sivaramakrishnan, and Suresh Jagannathan.
\newblock Composable asynchronous events.
\newblock In {\em Proceedings of the 32Nd ACM SIGPLAN Conf. on Programming
  Language Design and Implementation}, PLDI '11, pages 628--639, New York, NY,
  USA, 2011. ACM.

\end{thebibliography}




\end{document}